# Exact Solution of Spherical Harmonic Potential Coupled With E and M Fields

Tunde Joseph Taiwo
*Department of Physics, United Arab Emirate University, UAE*

**Abstract**: Using the technique of tridiagonal representation approach; for the first time, we extend this method to study quantum systems with literally perturbed Hamiltonians. Specifically, we consider a quantum system in a 3D spherical oscillator symmetric potential function coupled with an electric and a magnetic field. We obtain the energy spectrum and wavefunction in the two cases.

## 1. Introduction

Since the inception of nonrelativistic quantum mechanics, study has been primarily focused on precisely solvable potentials. The advantages gained from obtaining the corresponding properties of the underlying physical systems, such as eigenvalues, eigenfunctions, scattering phase shift, etc., form the basis of these. Additionally, these possibilities are positive effective physical system models or approximations. Various scientists have made attempts to categorize precisely solvable systems.

The Schrodinger problem had been successfully solved using the Tridiagonal Representation Approach. Here, the restriction on the diagonal matrix representation of Hamiltonian operators (wave operators) is loosened in this case, but tridiagonal and symmetric properties of the matrices are necessary. As a result, the solution space of the wave operator in a basis set, which need only be complete and square integrable with respect to a suitable configuration space measure, are increased; and basis set need not be a part of the Hamiltonian's eigenstates. By employing the TRA, we have been able to obtain novel solvable potential functions, including ones, which are not solvable or known in classical quantum mechanics. For further detailed understanding of this technique, readers are advised to read papers [1-4].

In this approach, the wave equation is replaced by the three term recursion relation of the orthogonal polynomials $\{P_n^\mu(\varepsilon), Q_n^\mu(\varepsilon)\}$. The action of the Hamiltonian operator on the basis elements goes as

$$H\left|\phi_n^\lambda\right\rangle = a_n\left|\phi_n^\lambda\right\rangle + b_{n-1}\left|\phi_{n-1}^\lambda\right\rangle + b_n\left|\phi_{n+1}^\lambda\right\rangle \tag{1}$$

where the constants $\{a_n, b_n\}$ depend on the physical parameters $\{\mu\}$ contained in $H$ which is written in the conventional format as the sum of the kinetic energy operator and the potential function; and $\left|\phi_n^\lambda\right\rangle$ is the basis function given in terms of the classical orthogonal polynomials. Using $H$ on the wave function expansion defined as

$$\left|\psi(x,\varepsilon)\right\rangle = \sqrt{\rho^\mu(\varepsilon)} \sum_{n=0}^{\infty} P_n^\mu(\varepsilon)\left|\phi_n^\lambda(x)\right\rangle \tag{2}$$

together with equation $H|\psi\rangle = E|\psi\rangle$, we obtain

$$EP_n^\mu(\varepsilon) = a_n P_n^\mu(\varepsilon) + b_{n-1} P_n^\mu(\varepsilon) + b_n P_{n+1}^\mu(\varepsilon) \tag{3}$$

where $E = E(\mu, \varepsilon)$. Therefore, the wave equation becomes equivalent to the three term recursion relation satisfied by the energy polynomials $\{P_n^\mu(\varepsilon)\}$. Of course solution of equation (3) can be obtained by comparing this recursion relation to recursion relation of well-known orthogonal polynomials.

In this present work we use the Tridiagonal representation approach, primarily based on the use of orthogonal polynomials, to literally perturbed Hamiltonians to obtain an analytic solution of the quantum systems. In section 2, we considered a quantum particle in 3D spherical harmonic oscillator potential function subjected to an electric field. This potential is mostly used to give a local model for certain attractive interactions, for example it is used to describe the local interaction of quarks. Infact, most attractive potentials are expanded near their stable points by the spherical harmonic oscillator potential. In section 3, we also consider similar quantum system subjected to a magnetic field. In both cases we obtain the energy spectrum and wavefunction. In section 4, conclusion and outlook, we discuss the further possible area of using the TRA in handling such non spherical central potential functions in a perturbed Hamiltonians under different circumstances.

## 2. The Spherical oscillator coupled with Electric field

We consider a particle of charge $q$ and mass $m$ in 3D spherical harmonic oscillator potential function of frequency $\omega$ subjected to an electric field $\zeta$. Here, we will use the Laguerre basis element (Appendix A) in the radial time independent Schrodinger wave equation.

$$(\hat{H} - E)|\psi\rangle = \left[ -\frac{\hbar^2}{2m}\frac{d^2}{dr^2} + \frac{\ell(\ell+1)}{2r^2} + \frac{1}{2}\omega^4 r^2 + q\zeta r^2 - E \right]|\psi\rangle = 0 \tag{5}$$

In conventional units $\hbar = m = 1$, equation (5) can be written as

$$(H - E)|\psi\rangle = -\frac{1}{2}\left[ (y')^2 \frac{d^2}{dy^2} + y''\frac{d}{dy} + \frac{\ell(\ell+1)}{r^2} - 2V(y) + 2E \right]|\psi\rangle \tag{6}$$

since $\frac{d}{dr} = y'\frac{d}{dy}$, $\frac{d^2}{dr^2} = (y')^2\frac{d^2}{dy^2} + y''\frac{d}{dy}$ and $V(y)$ is the potential function in the space transformation. With space configuration transformation $y = (\lambda r)^2$, equation (5) becomes

$$-\frac{1}{2\lambda^2}(H - E)|\psi\rangle = \left[ y\frac{d^2}{dy^2} + \frac{1}{2}\frac{d}{dy} - \frac{\ell(\ell+1)/4}{y} - \frac{1}{2\lambda^2}\{V(y) - E\} \right]|\psi\rangle \tag{7}$$

where $V(y) = \left(\frac{1}{2}\frac{\omega^4}{\lambda^2} + q\frac{\zeta}{\lambda^2}\right) y = \eta y$. By equation (2), the wave operator is expected to be

$$-\frac{1}{2\lambda^2}\sum_{n=0}^{\infty} P_n(\varepsilon)\langle\phi_m(y)|(H - E)|\phi_n(y)\rangle = -\frac{1}{2\lambda^2}\sum_{n=0}^{\infty} P_n(\varepsilon)\langle\phi_m(y)|\left[ y\frac{d^2}{dy^2} + \frac{1}{2}\frac{d}{dy} - \frac{\ell(\ell+1)/4}{y} - \frac{1}{2\lambda^2}\{V(y) - E\} \right]|\phi_n(y)\rangle \tag{8}$$

and the wave operator matrix elements is

$$-\frac{1}{2\lambda^2}J_{mn} = -\frac{1}{2\lambda^2}\langle\phi_m|(H-E)|\phi_n(y)\rangle = -\frac{1}{2\lambda^2}\langle\phi_m(y)|\left[y\frac{d^2}{dy^2}+\frac{1}{2}\frac{d}{dy}-\frac{\ell(\ell+1)/4}{y}-\frac{1}{2\lambda^2}\{V(y)-E\}\right]|\phi_n(y)\rangle \qquad (9)$$

with integral measure of $\int_0^\infty \phi_m(r)F[r]\phi_n dr = \int_0^\infty \phi_m(y)F[y]\phi_n(y)\frac{dy}{2\lambda\sqrt{y}}$ where

$F[y] = y\frac{d^2}{dy^2}+\frac{1}{2}\frac{d}{dy}-\frac{\ell(\ell+1)/4}{y}-\frac{1}{2\lambda^2}\{V(y)-E\}$. Since the basis element is $|\phi_n(r)\rangle = A_n y^\alpha e^{-\beta y}L_n^v(y)$, with

$A_n = \sqrt{\frac{2\lambda\Gamma(n+1)}{\Gamma(n+v+1)}}$, then normalization condition is

$$\int_0^\infty \phi_n(r)\phi_m(r)dr = A_n A_m \int_0^\infty y^{2\alpha}e^{-2\beta}L_n^v(y)L_m^v(y)\cdot\frac{dy}{2\lambda\sqrt{y}} = \frac{\Gamma(n+v+1)}{\Gamma(n+1)}\delta_{nm} \qquad (10)$$

Since

$$\frac{d}{dy}|\phi_n\rangle = A_n y^\alpha e^{-\beta y}\left[\frac{d}{dy}+\frac{\alpha}{y}-\beta\right]L_n^v(y) \qquad (11a)$$

and

$$\frac{d^2}{dy^2}|\phi_n\rangle = A_n y^\alpha e^{-\beta y}\left[\frac{d^2}{dy^2}+\left(\frac{2\alpha}{y}-2\beta\right)\frac{d}{dy}-\frac{\alpha}{y^2}+\left(\frac{\alpha}{y}-\beta\right)^2\right]L_n^v(y) \qquad (11b)$$

Using these with equation (A4) and (A5) in appendix, we have

$$-\frac{1}{2\lambda^2}J_{mn} = \frac{1}{2\lambda}A_n A_m \int_0^\infty y^{2\alpha-1/2}e^{-2\beta y}\left\{\begin{array}{l}\left[\left(-v-\frac{1}{2}+2\alpha\right)\frac{n}{y}-(2\beta y-y)\frac{n}{y}-n+\frac{\alpha(\alpha-1)}{y}-\frac{\ell(\ell+1)/4}{y}-2\alpha\beta+y\beta^2-\frac{1}{2\lambda^2}(V(y)-E)\right]L_n^v(y)\\ -\frac{(n+v)}{y}\left(-v-\frac{1}{2}+2\alpha\right)L_{n-1}^v(y)+\frac{(n+v)}{y}(2\beta y-y)L_{n-1}^v(y)\end{array}\right\}L_m^v(y)dy \qquad (12)$$

since we require triadiagonality and symmetry, the parameters of the basis element will take $\alpha = \frac{v}{2}+\frac{1}{4}, \beta = \frac{1}{2}$ and $v = \ell + \frac{1}{2}$, therefore equation (12) becomes

$$\frac{1}{2\lambda^2}J_{mn} = \frac{1}{2\lambda}A_n A_m \int_0^\infty y^{2\alpha-1/2}e^{-2\beta y}\left\{\left[\frac{2n+v+1}{2}-\frac{y}{4}+\frac{1}{2\lambda^2}(V(y)-E)\right]L_n^v(y)\right\}L_m^v(y)dy \qquad (13)$$

This can be rearranged as

$$\frac{1}{2\lambda^2}J_{mn} = \frac{1}{2\lambda}A_n A_m \int_0^\infty y^{2\alpha-1/2}e^{-2\beta y}\left\{\left[\frac{2n+v+1}{2}+y\xi-\frac{E}{2\lambda^2}\right]L_n^v(y)\right\}L_m^v(y)dy \qquad (14)$$

where $\xi = \left(\frac{\omega^4}{4\lambda^4}+\frac{q\zeta}{2\lambda^4}-\frac{1}{4}\right)$. Further calculations in (14), we have

$$\frac{1}{2\lambda^2}J_{mn} = \left\{\left[\left(\frac{1}{2}+\xi\right)(2n+v+1)-\frac{E}{2\lambda^2}\right]\delta_{nm}-\xi\sqrt{n(n+v)}\delta_{n,m+1}-\xi\sqrt{(n+1)(n+v+1)}\delta_{n,m-1}\right\} \qquad (15)$$

And by equation (8) we have,

$$\frac{1}{\xi}\left[\left(\frac{1}{2}+\xi\right)(2n+v+1)-\frac{E}{2\lambda^2}\right]P_n = \sqrt{n(n+v)}P_{n-1}+\sqrt{(n+1)(n+v+1)}P_{n+1} \tag{16}$$

The discrete energy spectrum of the Hamiltonian can be obtained from equation (15) by diagonalizing; with the requirement that $\xi = 0$ and $\lambda^2 = \omega^2$, therefore

$$E_n = \lambda^2(2n+v+1) = \omega^2(2n+v+1) \tag{17}$$

where $v = \ell + 1/2$ and $n = 0,1,2,...,$. We can conclude that the electric field has no effect on the energy spectrum in the sense that energy spectrum is not splitted due to the presence of the electric field. If we use the recurrence relation for the Meixner Pollaczek polynomial

$$2y\sin\theta f_n^\mu(y,\theta) = -\left[(2n+2\mu)\cos\theta\right]f_n^\mu(y,\theta)+\sqrt{n(n+2\mu-1)}f_{n-1}^\mu(y,\theta)+\sqrt{(n+1)(n+v+1)}f_{n+1}^\mu(y,\theta) \tag{18}$$

Of course we can get the hyperbolic Meixner Pollaczek [11] by $\theta \to i\theta$, then $\sin\theta = i\sinh\theta$, and $\cos\theta = \cosh\theta$ which gives

$$i2y\sin\theta f_n^\mu(y,\theta) = -\left[(2n+2\mu)\cosh\theta\right]f_n^\mu(y,\theta)+\sqrt{n(n+2\mu-1)}f_{n-1}^\mu(y,\theta)+\sqrt{(n+1)(n+v+1)}f_{n+1}^\mu(y,\theta) \tag{19}$$

When compared to (16), of course gives $2\mu = v+1 = \ell+3/2$. This makes the polynomial in wavefunction (2) as

$$P_n^\mu(y,\theta) = \sqrt{\frac{(2\mu)_n}{n!}}e^{-n\theta}\,_2F_1\left(\begin{array}{c}-n,\mu+iy\\2\mu\end{array}\bigg|1-e^{-2\theta}\right) \tag{20}$$

Of course the wavefunction is affected by the electric field because $\cosh\theta = \frac{1}{\xi}\left(\frac{1}{2}+\xi\right) = \frac{\omega^4}{2q\xi}$.

### 3. The Spherical oscillator coupled with magnetic field

In [8], the Hamiltonian of a particle of mass $m$ and charge $q$ moving in a central potential $V(r)$ under the influence of a uniform magnetic field $\vec{B}$.

$$\hat{H} = \frac{1}{2m}\vec{p}^2 + V(r) - \frac{q}{2mc}\vec{B}.\hat{\vec{L}} + \frac{q^2}{8mc^2}\left[B^2r^2-(\vec{B}.\vec{r})^2\right] \tag{20}$$

where $c$ is the speed of light, $\hat{\vec{L}}$ is orbital angular momentum operator of the particle. And the spin orbital magnetic dipole momentum is neglected. Now the radial Schrodinger equation will take the form

$$(\hat{H}-E)|\psi(r)\rangle = \left[-\frac{\hbar^2}{2m}\frac{d^2}{dr^2}+\frac{\ell(\ell+1)}{2mr^2}+V(r)-\frac{q}{2mc}\vec{B}.\hat{\vec{L}}+\frac{q^2}{8mc^2}\left[B^2r^2-(\vec{B}.\vec{r})^2\right]-E\right]|\psi(r)\rangle = 0 \tag{21}$$

The magnetic field made the Hamiltonian a perturbed one. For simplicity sake, we assumed the magnetic field is along the $z$ axis. Therefore, $\frac{q}{2mc}\vec{B}.\hat{\vec{L}} = \frac{q}{2mc}B\hat{L}_z$ which is known as the paramagnetic component $\hat{H}$. Also $\frac{q^2}{8mc^2}\left[B^2r^2-(\vec{B}.\vec{r})^2\right] = \frac{q^2}{8mc^2}\left[B^2r^2-B^2r_z^2\right] = \frac{q^2B^2}{8mc^2}\left[r^2-r_z^2\right] = \frac{q^2B^2}{8mc^2}\left[r_x^2+r_y^2\right]$ since $r^2 = r_x^2+r_y^2+r_z^2$. This is also known as the diamagnetic term of $\hat{H}$. Suppose we are in interested in proving the **Normal Zeeman effect** usually experience in the hydrogen atom, of course we could neglect the diamagnetic term where $V(r) = -\frac{1}{4\pi\varepsilon_0}\frac{e^2}{r}$, $q = -e$, and $\langle r_x^2 + r_y^2 \rangle \approx a_0^2$ (Bohr radius). The TRA can also be used to show this result. On a more generic case we tend to see how the energy spectrum and wavefunction of equation (21) would be when the potential is spherical harmonic potential and what will be the aftermath. Now we assumed that $\frac{q^2 B^2}{8mc^2}[r_x^2 + r_y^2] = \frac{q^2 B^2}{8mc^2} 2r_x^2 \approx \frac{q^2 B^2}{4mc^2} r^2$. Also since the $\hat{H}$ and $\hat{L}_z$ commute, in view of TRA where $\psi(r) = \sum_{n=0}^{\infty} f_n(\varepsilon)\phi_n(r)$, we have

$$\int_0^\infty \phi_m(r)\left(\frac{q}{2mc} B\hat{L}_z\right)\phi_n(r)dr = \frac{q}{2mc} B\mu\hbar \int_0^\infty \phi_m(r)\phi_n(r)dr = \frac{q}{2mc} B\mu\hbar \delta_{mn}, \tag{22}$$

where $\mu$ is the azimuthal quantum number. In the presence of a spherical harmonic potential function equation (21) becomes (in conventional units $\hbar = m = 1$)

$$(\hat{H} - E)|\psi\rangle = \left[-\frac{1}{2}\frac{d^2}{dr^2} + \frac{\ell(\ell+1)}{2r^2} + V(r) - \frac{q}{2c} B\hat{L}_z - E\right]|\psi\rangle = 0 \tag{23}$$

where $V(r) = \left(\frac{1}{2}\omega^4 + \frac{q^2 B^2}{4c^2}\right)r^2$. With same space configuration transformation $y = (\lambda r)^2$, equation (23) becomes

$$-\frac{1}{2\lambda^2}(H - E)|\psi\rangle = \left[y\frac{d^2}{dy^2} + \frac{1}{2}\frac{d}{dy} - \frac{\ell(\ell+1)/4}{y} - \frac{1}{2\lambda^2}\left\{V(y) - \frac{q}{2c} B\hat{L}_z - E\right\}\right]|\psi\rangle \tag{24}$$

and $V(y) = \left(\frac{\omega^4}{2\lambda^2} + \frac{q^2 B^2}{4\lambda^2 c^2}\right)y = \eta y$. Following same process as in section 2, from equation (12), with the requirement for triadiagonality and symmetry; the parameters of the basis element will take $\alpha = \frac{v}{2} + \frac{1}{4}, \beta = \frac{1}{2}$ and $v = \ell + \frac{1}{2}$, we can deduce that the wave operator matrix elements is given by

$$\frac{1}{2\lambda^2} J_{mn} = \frac{1}{2\lambda} A_n A_m \int_0^\infty y^{2\alpha - 1/2} e^{-2\beta y} \left\{\left[\frac{2n+v+1}{2} - \frac{y}{4} + \frac{1}{2\lambda^2}\left(V(y) - \frac{q}{2c} B\hat{L}_z - E\right)\right] L_n^v(y)\right\} L_m^v(y) dy \tag{25}$$

Rearranging,

$$\frac{1}{2\lambda^2} J_{mn} = \frac{1}{2\lambda} A_n A_m \int_0^\infty y^{2\alpha - 1/2} e^{-2\beta y} \left\{\left[\frac{2n+v+1}{2} + y\xi_1 - \frac{E}{2\lambda^2} - \frac{qB}{4\lambda^2 c}\hat{L}_z\right] L_n^v(y)\right\} L_m^v(y) dy \tag{26}$$

where $\xi_1 = \frac{\omega^4}{4\lambda^4} + \frac{q^2 B^2}{8\lambda^4 c^2} - \frac{1}{4}$. Further calculations yield

$$\frac{1}{2\lambda^2} J_{mn} = \left\{\left[\left(\frac{1}{2} + \xi_1\right)(2n+v+1) - \frac{E}{2\lambda^2} - \frac{qB}{4\lambda^2 c}\mu\right]\delta_{nm} - \xi_1\sqrt{n(n+v)}\delta_{n,m+1} - \xi_1\sqrt{(n+1)(n+v+1)}\delta_{n,m-1}\right\} \tag{27}$$

Of course this same as

$$\frac{1}{\xi_1}\left[\left(\frac{1}{2} + \xi_1\right)(2n+v+1) - \frac{E}{2\lambda^2} - \frac{qB}{4\lambda^2 c}\mu\right] P_n = \sqrt{n(n+v)} P_{n-1} + \sqrt{(n+1)(n+v+1)} P_{n+1} \tag{28}$$

The discrete energy spectrum of the Hamiltonian can be obtained from equation (27) by diagonalizing; with the requirement that $\xi_1 = 0$ and $\lambda^2 = \omega^2$, therefore

$$E_n = \omega^2(2n+v+1) - \frac{qB}{2c}\mu \tag{29}$$

where $v = \ell + 1/2$, $\mu$ is the azimuthal quantum number, and $n = 0, 1, 2,...$, This means that the whole oscillator spectrum is shifted up/down by a constant. Interestingly, equation (29) is similar to **the normal Zeeman effect in Hydrogen atom when the Coulomb potential is used**. Of course if we take $q \to -e$ and retain conventional units $m$ and $\hbar$, we have

$$E_{n\ell\mu} = \omega^2(2n+v+1) + \mu\hbar\omega_\ell \tag{30}$$

where $\omega_\ell = \frac{eB}{2mc}$ is Lamar frequency. Of course the wave function can be obtained similar to equation (20). And we also have $\cosh\theta = \frac{1}{\xi_1}\left(\frac{1}{2} + \xi_1\right) = \frac{4\omega^4 c^2}{q^2 B^2}$.

### 4. Conclusion and outlook

In this paper, we obtain the solution of the Schrodinger wave equation when the Hamiltonian is composed of the spherical oscillator coupled with an electric and magnetic field by using the tridiagonal representational approach. Interestingly, we are able analytically get the energy spectrum; and a similar result to the normal Zeeman Effect; and also the wavefunction for the system was obtained and shows that the energy spectrum of the spherical oscillator is affected by the magnetic field. Here, equation (29) is opened to physical parameters depending on the physical insight needed. We tend to follow future works by applying the TRA to solve quantum systems with non-spherical central potential function in a perturbed Hamiltonians which may or may not be solvable in conventional quantum mechanics then we compare our results. Such potentials includes the improved ring shaped non spherical harmonic oscillators

$$V(r,\theta) = \frac{1}{2}m\omega^2 r^2 + \frac{1}{2}\frac{\hbar^2\eta}{mr^2} + \frac{\hbar^2}{2mr^2}\left(\frac{\beta\sin^2\theta + \gamma\cos^2\theta + \lambda}{\sin\theta\cos\theta}\right)^2 \tag{31}$$

and the improved ring shaped coulomb potential given as

$$V(r,\theta) = \frac{-\beta}{r} + \frac{\hbar^2}{2mr^2}\left(\frac{\beta\sin^2\theta + \gamma\cos^2\theta + \lambda}{\sin\theta\cos\theta}\right)^2 \tag{32}$$

where $\beta$, $\gamma$ and $\lambda$ are the adjustable potential parameters. Special cases of these potential is the Ring shape non spherical harmonic potential

$$V(r,\theta) = \frac{1}{2}m\omega^2 r^2 + \frac{1}{2}\frac{\hbar^2\alpha}{mr^2} + \frac{\hbar^2}{2mr^2\sin^2\theta} \tag{33}$$

and Harmonic novel angle dependent potential

$$V(r,\theta) = \frac{1}{2}m\omega^2 r^2 + \frac{\hbar^2}{2mr^2}\left(\frac{\eta + A\cos^2\theta + B\cos^4\theta}{\sin^2\theta\cos^2\theta}\right) \tag{34}$$

These and many others are worth looking using TRA because the solution space of the wave operator is wider than the conventional form in quantum mechanics. Hence more interesting results will be obtained.

### Appendix A

**Laguerre Basis:** This basis element is given as

$$|\phi_n(r)\rangle = A_n y^\alpha e^{-\beta y} L_n^v(y) \tag{A1}$$

Where $\alpha$ and $v$ are real parameters with $v > -1$ and $\alpha \geq 0$ to ensure convergence of the Laguerre polynomial $L_n^v(y)$ and compatibility with boundary conditions when the new variable $y$ span semi-infinite interval $[0,\infty]$. Also $A_n = \sqrt{\lambda \Gamma(n+1)/\Gamma(n+v+1)}$, in choosing $A_n$ sometime insight is required based on the derivatives of the configuration space transformation coordinate. The Laguerre polynomial $L_n^v(y)$ satisfies the following properties

$$y L_n^v(y) = (2n+v+1) L_n^v(y) - (n+v) L_{n-1}^v(y) - (n+1) L_{n+1}^v(y) \tag{A2}$$

$$L_n^v(y) = \frac{\Gamma(n+v+1)}{\Gamma(n+1)\Gamma(v+1)} {}_1F_1(-n; v+1, y) \tag{A3}$$

$$\left[ y \frac{d^2}{dy^2} + (v+1-y) \frac{d}{dy} + n \right] L_n^v(y) = 0 \tag{A4}$$

$$y \frac{d}{dy} L_n^v(y) = n L_n^v(y) - (n+v) L_{n-1}^v(y) \tag{A5}$$

$$\int_0^\infty y^v e^{-y} L_n^v(y) L_m^v(y) dy = \frac{\Gamma(n+v+1)}{\Gamma(n+1)} \delta_{nm} \tag{A6}$$

**References:**
[1] A.D. Alhaidari, "*An extended class of 2 L - series solution of the wave equation*," Ann. Phys **317,** 152 (2005)

[2] A.D. Alhaidari, *Solution of the nonrelativistic wave equation using the tridiagonal representation approach*, J. Math. Phys. **58,** 072104 (2017)

[3] A.D. Alhaidari, "Analytic *solution of the wave equation for an electron in the field of a molecule with an electric dipole moment*, " Ann. Phys, **323**, 1709 (2008)

[4] A.D. Alhaidari and T.J Taiwo,"*Four – parameter potential box with inverse square singular boundaries*" .Eur. Phys. J. Plus 133 (2018) 115

[5] T .J .Taiwo, "*Four parameters Potential Function with Negative Energy Bound State*". Inf. Sci.Lett.8, No 1, 25 - 31 (2019)

[6] T.J. Taiwo, E.O. Oghre and A. N. Njah, "Solution of Schrodinger equation using Tridiagonal representation approach in nonrelativistic quantum mechanics".Transactions Vol 8 Jan 2019. Journal of the Nigerian Association of Mathematical physics.

[8]Nouredine Zettilli, *Qauntum mechanics – Concepts and Applications*, John Wiley and Sons, Ltd, 2001

[7] A.D. Alhaidari, "*Scattering and bound states for a class of non –central potentials*", J. Phys. A: Math. Theor.**38**. 3409 (2005)

[8] A.D. Alhaidari, "*A class of singular logarithmic potentials in a box with different skin thickness and wall Interactions,*" Phys. Scr. **82**, 065008 (2010)

[9] A.D. Alhaidari, "*Exact solutions of Dirac and Schrodinger equations for a large class of power- law potentials at zero energy,*" Int. J. Mod. Phys. A **17,** 4551 (2002)

[10] P.C. Ojihe, "*SO(2,1) Lie algebra, the Jacobi matrix and scattering states of the Morse oscillator*," J. Phys. A: Math. Gen **21**, 875 (1988)

[11] W. Magnus, F. Oberhettinger, R.P. Soni, *Formulas and Theorems for the Special Functions of Mathematical Physics,* Springer, New York, 1966;

T.S. Chihara, *An Introduction to Orthogonal Polynomials*, Gordon and Breach, New York, 1978;

G. Szegö, *Orthogonal Polynomials*, fourth ed., Am. Math. Soc, Providence, RI, 1997;

R. Askey, M. Ismail*, Recurrence relations, continued fractions and orthogonal polynomials*, Memoirs of the Am. Math. Soc., Vol. 49 Nr. 300 (Am. Math. Soc., Providence, RI, 1984).

[12] A.D. Alhaidari, *"On the asymptotic solutions of the scattering problem*", J. Phys. A: Math. Theor. **41** 175201 (2008)

[13] R.W. Haymaker and L. Schlessinger, *The Pade Approximation in Theoretical Physics*, edited by G.A. Baker and J.L Gammel (Academic Press, New York, 1970)

[14] A.D. Alhaidari and H. Bahlouli, *Extending the class of solvable potentials I. The infinite potential well with a sinusoidal bottom,* J. Math. Phys. 49 (2008) 082102